\documentclass[journal]{IEEEtran}
\usepackage{amsmath,amsfonts}
\usepackage{amssymb} 
\usepackage{algorithmic}
\usepackage{algorithm}
\usepackage{array}
\usepackage[caption=false,font=normalsize,labelfont=sf,textfont=sf]{subfig}
\usepackage{textcomp}
\usepackage{stfloats}
\usepackage{url}
\usepackage{verbatim}
\usepackage{graphicx}
\usepackage{cite}
\usepackage{amsthm}
\theoremstyle{plain}
\newtheorem{theorem}{Theorem}
\usepackage{mathtools}  

\usepackage{booktabs}
\usepackage{siunitx}
\allowdisplaybreaks 
\usepackage{booktabs}
\usepackage{hhline}
\usepackage[table]{xcolor}
\usepackage{diagbox}

\newcommand{\NA}{}

\usepackage{amsmath,amssymb,mathtools}

\definecolor{darkgreen}{rgb}{0.0, 0.673, 0.51}
\definecolor{darkcyan}{rgb}{0.0, 0.35, 0.55}

\usepackage[colorlinks=true, linkcolor=black, citecolor=black, urlcolor=black, pdfborder={0 0 0}]{hyperref}  

\newtheorem{remark}{Remark}
\newcommand{\stirlingii}[2]{\genfrac{\{}{\}}{0pt}{}{#1}{#2}}
\newcommand{\eqa}{\mathrel{\stackrel{\scriptstyle\text{(a)}}{=}}}
\newcommand{\eqb}{\mathrel{\stackrel{\scriptstyle\text{(b)}}{=}}}

\newcommand{\orcidicon}[1]{\href{https://orcid.org/#1}{\raisebox{0.4ex}{\includegraphics[height=1.6ex]{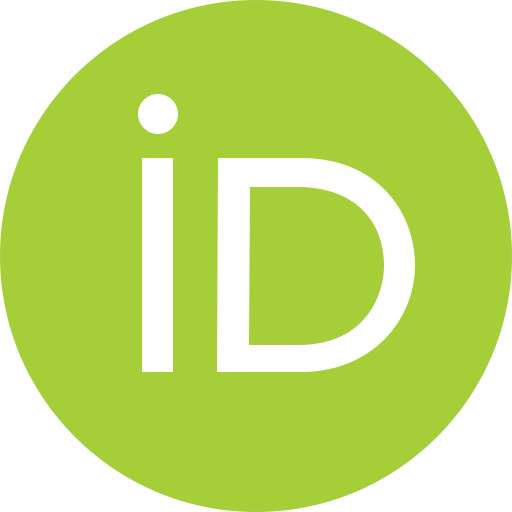}}}}

\usepackage{fancyhdr}

\fancypagestyle{firstpage}{%
  \fancyhf{} 
  
  \fancyhead[L]{\scriptsize \copyright~ 2026 IEEE. Personal use of this material is permitted.  Permission from IEEE must be obtained for all other uses, in any current or future media, including reprinting/republishing this material for advertising or promotional purposes, creating new collective works, for resale or redistribution to servers or lists, or reuse of any copyrighted component of this work in other works. 
  DOI: \href{https://doi.org/10.1109/LSP.2026.3661843}{\textcolor{darkcyan}{10.1109/LSP.2026.3661843}}}%
}
\begin{document}
\makeatletter
\typeout{IEEEtran text font size = \f@size pt}
\makeatother

\title{Efficient Computation of Time-Index Powered Weighted Sums Using Cascaded Accumulators}

\author{Deijany Rodriguez Linares\orcidicon{0009-0004-1846-9496}, \textit{Graduate Student Member, IEEE}, Oksana Moryakova\orcidicon{0009-0001-6464-5452}, \textit{Graduate Student Member, IEEE}, and H\aa kan Johansson\orcidicon{0000-0001-6329-9132}, \textit{Senior Member, IEEE}%
\thanks{The authors are with the Department of Electrical Engineering, Linköping University, 58183 Linköping, Sweden. Email: \{deijany.rodriguez.linares, oksana.moryakova, hakan.johansson\}@liu.se.}
\thanks{This work was funded by ELLIIT and Sweden's innovation agency.}}

\maketitle
\thispagestyle{firstpage} 
\begin{abstract}
\sloppy
This letter presents a novel approach for \mbox{efficiently} computing time-index powered weighted sums of the form $\sum_{n=0}^{N-1} n^{K} v[n]$ using cascaded accumulators. 
Traditional direct computation requires $K{\times}N$ general multiplications, which become prohibitive for large $N$, while alternative strategies based on lookup tables or signal reversal require storing entire data blocks. By exploiting accumulator properties, the proposed method eliminates the need for such storage and reduces the multiplicative cost to only $K{+}1$ constant multiplications, enabling efficient real-time implementation. The approach is particularly useful when such sums need to be efficiently computed in sample-by-sample processing systems.

\end{abstract}
\begin{IEEEkeywords}
Accumulators, addition-chain exponentiation, binomial coefficients, Stirling numbers, time-index powered weighted sums.
\end{IEEEkeywords}

\section{Introduction}

\IEEEPARstart{W}{e} define time-index powered weighted sums as
\begin{equation}
S = \sum_{n=0}^{N-1} n^{K} v[n], \quad K, N \in \mathbb{Z}_{\geq 0}, \label{eq:S_k}
\end{equation}
where $v[n]$ is an arbitrary sequence (discrete-time signal with $n$ being the time index) and $K$ is the power. Such sums arise in signal and image processing, and can be viewed as a particular case of moment computations. Specifically, the one-dimensional discrete (or geometric) moments have been utilized in image analysis~\cite{hatamian86, Zhou2000}. They also arise in synchronization tasks, such as sampling frequency and time-offset estimation, where the cases \mbox{$K=0,1,2$} are of particular importance~\cite{DOH_SF0_DSP}. However, a straightforward computation of~\eqref{eq:S_k} requires $K{\times}N$ general multiplications and $N{-}1$ additions, which becomes a limiting factor as $N$ increases, especially in resource-constrained systems or applications with low-power requirements.

Given that multiplications are typically more costly than additions in terms of power, area, and latency~\cite{Kumm2023, parhi2007vlsi, koren2001computer, Aksoy_2014}, prior efforts have focused on reducing this overhead. Strategies include optimal addition-chain exponentiation to minimize the cost of generating $n^K$~\cite{hatem2021}, and lookup tables (LUTs) to precompute powers at the expense of area~\cite{parhi2007vlsi}. However, both strategies still require the $N$ multiplications (general in the case of addition-chains, constant\footnote{We distinguish between general and constant multiplications. General multiplications require full multipliers with arbitrary inputs, while constant multiplications involve one fixed operand and can be implemented efficiently using adders and shifters~\cite{Voronenko2007MultiplierlessMC, DOH_SF0_DSP}.} for LUTs) \mbox{by $v[n]$}.

In this letter, we propose a method to compute~\eqref{eq:S_k} using a cascade of $K{+}1$ accumulators that process data in real time on a sample-by-sample basis, i.e., in forward time without block buffering or time reversal. This eliminates the need to store the entire input sequence, requiring only $K{+}1$ storage registers (one per accumulator), regardless of the sequence length $N$, followed by a final step with $K{+}1$ constant multiplications by precomputed coefficients. To the best of the authors' knowledge, this is the first method for computing the sum in real time, sample-by-sample, using cascaded accumulators, offering significant advantages for embedded and low-power systems where real-time processing and memory efficiency are critical~\cite{Bjornson24}.

It is noted that accumulator-based structures have previously been studied in~\cite{Zhou2000} to compute~\eqref{eq:S_k}. However, the approach in~\cite{Zhou2000} reverses the input signal (i.e., $n \mapsto N{-}1{-}n$), which requires buffering an entire block of $N$ samples (i.e., reverse-order evaluation of the same sum) and is therefore not sample-by-sample processing in forward time. Other accumulator-based methods have considered sums with reversed time-index powers $(N{-}1{-}n)^K$~\cite[Sec.~II, eq.~(3)]{hatamian86}, as well as efficient filter implementations~\cite{Saramaki95}. However, these techniques are not applicable to the forward time-index powered sum in~\eqref{eq:S_k}, as they address different polynomial indexings and filter realizations respectively, and do not target the causal forward-time computation considered here. Our work specifically addresses this challenge by introducing a new architecture tailored for this sum, combining the computational efficiency of cascaded accumulators with the practicality of real-time sample-by-sample processing.

Following this introduction, Section~\ref{sec:accum} derives the cascaded accumulators outputs. Section~\ref{sec:problem} presents the proposed method whereas Section~\ref{sec:results} gives a complexity analysis and implementation results, and Section~\ref{sec:conclusion} concludes the letter.

\it Notation and Definitions: \rm We denote the binomial coefficient for a non-negative integer~\!by
\begin{equation}
  \binom{n}{k} {\triangleq} \hspace{-0 pt} \frac{n(n\hspace{-0 pt}{-}\hspace{-0 pt}1){\cdots}(n\hspace{-0 pt}{-}\hspace{-0 pt}k\hspace{-0 pt}+\hspace{-1 pt}1)}{k!} {=} \frac{n!}{k!(n {-} k)!}, \quad n, k {\in} \mathbb{Z}_{\geq 0} , \label{eq:binom_def}
\end{equation}
with $n! = n(n - 1)(n - 2)\cdots 1$ being the factorial of a non-negative integer $n$ (by convention, $0! = 1$). 

Further the $j$-th rising factorial (or Pochhammer symbol) is defined by\begin{equation}
  x^{\overline{j}} \triangleq x(x+1)\cdots(x+j-1), \quad x \in \mathbb{R},\ j \in \mathbb{Z}_{\geq 0}.
  \label{eq:rising}
\end{equation}
For an integer $x \geq 1$, this can alternatively be expressed as
\begin{equation}
  x^{\overline{j}} \eqa \frac{(x+j-1)!}{(x-1)!} = \frac{j!}{j!}\frac{(x+j-1)!}{(x-1)!} \eqb j!\binom{x{+}j{-}1}{j},
  \label{eq:rising_ref}
\end{equation}
where (a) follows from the identity in~\mbox{\cite[Sec.~1.4, eq.~(4d)]{Neumann2016AdvancedCT}} and (b) from applying \eqref{eq:binom_def}. 

The Stirling numbers of the second-kind,\footnote{Stirling numbers of the second kind, $\stirlingii{n}{k}$, count the number of ways to partition a set of $n$ elements into $k$ nonempty subsets.} defined by 
\begin{equation}
  \stirlingii{n}{k} = \frac{1}{k!} \sum_{i=0}^k (-1)^i \binom{k}{i} (k-i)^n,
  \label{eq:stirling_second}
\end{equation}
satisfy the identity
\begin{equation}
  x^m = \sum_{j=0}^{m} \stirlingii{m}{j} (-1)^{m-j} x^{\overline{j}}, \label{eq:stirling_expansion}
\end{equation}
which expresses monomials in terms of rising factorials, as shown in~\cite[eq.~(6.12)]{concrete_mathematics}.
\section{Cascaded Accumulator's Output} \label{sec:accum}
\begin{figure}[tbp]
  \centering
  \includegraphics[scale=0.87]{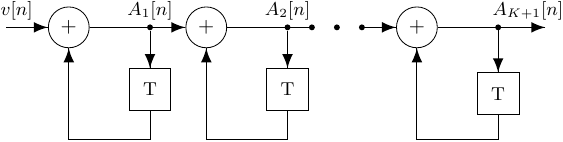 }
  \caption{$K{+}1$ cascaded accumulators.}
  \label{flo:cascade}
\end{figure}

Consider a cascade of $K\!+\!1$ discrete-time accumulators, as depicted in Fig.~\ref{flo:cascade}, where for an input $v[n]$, with \mbox{$n\!=\!0,1,\dots, N\!-\!1$}, the $k$-th accumulator computes the \mbox{recurrence}
\begin{equation}
  \hspace{-0 pt} A_k[n]\hspace{-3 pt} = \hspace{-3 pt}A_k[n\hspace{-1.0 pt}-\hspace{-1.0 pt}1] \hspace{-1 pt}+ \hspace{-1 pt}A_{k-1}[n], \hspace{0.0 pt}
  \begin{cases}
    \hspace{-1 pt}A_0[n] = v[n], \,\, n\ge0,\\
    \hspace{-1 pt}A_k[\hspace{-1 pt}-1\hspace{-1 pt}] = 0, \,\, k \hspace{-2 pt}\in \hspace{-1 pt} \mathbb{Z}[0, K\!+\!1].
  \end{cases}\hspace{-16.35 pt}
  \end{equation}

Each accumulator corresponds to a first-order IIR filter with transfer function $1/(1 - z^{-1})$. Thus, the transfer function from the input $v[n]$ to the $k$-th output $A_k[n]$ is
\begin{equation}
  H_k(z) 
    = \prod_{i=1}^k \frac{1}{1 - z^{-1}} 
    = \frac{1}{(1 - z^{-1})^k},
\end{equation}
with the corresponding impulse response given by its inverse \mbox{$z$-transform}, as
\begin{align}
  h_k[n] &= \underset{k \text{ times}}{\underbrace{u[n]*u[n]*\dots *u[n]}} 
  \eqa \binom{n + k - 1}{k - 1}u[n], 
  \label{eq:impulse}
\end{align}
where $*$ denotes convolution, $u[n]$ is the unit-step function \mbox{(i.e., $u[n]=1$ $ \forall n \geq 0$, and $0$ otherwise)} and $\binom{n+k-1}{k-1}$ is the binomial coefficient as defined in \eqref{eq:binom_def}. Step (a) follows from the \mbox{$z$-transform} pair given in \mbox{\cite[Sec.~13.4, Prop.~z14]{mathematics_handbook_beta}}.

Then, for an input $v[n]$, the output of the $k$-th accumulator, after $N$ samples, is given by the discrete-time convolution
\begin{align}
A_k{[N-1]} &= (v \ast h_k)[N-1] = \sum_{n=0}^{N-1} v[n]h_k[(N-1)-n] \nonumber \\
&\eqb \sum_{n=0}^{N-1} \binom{N - n + k - 2}{k - 1}v[n], \label{eq:accum} 
\end{align}
where (b) is obtained by applying \eqref{eq:impulse}.
\begin{figure}[tbp]
  \centering
  \includegraphics[scale=0.87]{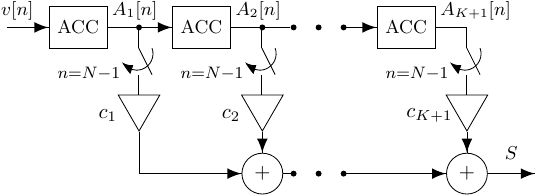 }
  \caption{Structure for computing time-index weighted sums via \mbox{$K{+}1$} cascaded accumulators ($\mathrm{ACC}$).}
  \label{flo:pipeline}
\end{figure}

The expression for $A_k[N-1]$ in~\eqref{eq:accum} reveals that the \mbox{$k$-th} accumulator computes a weighted sum of the input $v[n]$, where the weights are given by the binomial term $\binom{N - n + k - 2}{k - 1}$, a degree-$(k{-}1)$ polynomial in~$n$.\footnote{\label{foo:order}From~\eqref{eq:binom_def}, the numerator has $k{-}1$ terms, making it a degree-($k{-}1$) polynomial in $n$: $\binom{N{-}n{+}k{-}2}{k{-}1} {=} \tfrac{\prod_{i=0}^{k-2}(N{-}n{+}i)}{(k{-}1)!}$.}

\section{Proposed Computation of Time-Index Powered Weighted Sums via Cascaded Accumulators} \label{sec:problem}
Since the accumulator outputs given by \eqref{eq:accum}, for {$k=1, 2,\ldots, K{+}1$}, form a basis of polynomials of distinct degrees $0$ through $K$, and any degree-$K$ polynomial can be expressed as a linear combination of this basis,\footnote{See, e.g., the Newton basis or binomial basis for polynomials as discussed in \cite[Chap.~9]{humpherys2020foundations}.} it is natural to express the time-index powered weighted sum in~\eqref{eq:S_k} as a linear combination of the outputs $A_k[N{-}1]$. This leads to the representation depicted in Fig. \ref{flo:pipeline}, which is mathematically described by
\begin{align}
  S=\sum_{n=0}^{N-1} n^{K} v[n] = \sum_{k=1}^{K+1} c_k A_k[N-1], \quad K \in \mathbb{Z}_{\geq 0}, \label{eq:c_kA_k}
\end{align}
where each coefficient $c_k$ can be expressed as a function of $N$ in closed form, as will be proved later in this section. It is noted that the multiplications by $c_k$ are carried out only once after the accumulations, i.e., at sample $n=N{-}1$. This enables an efficient computation of the sum over $N$ samples using only\footnote{Typically, $K \ll N$ in practice. For example, in ~\cite{DOH_SF0_DSP}, $K=2$ whereas $N=256$.}
\begin{itemize}
  \item $K{+}1$ constant multiplications (one per accumulator),
  \item $K{+}1$ storage registers (one per accumulator), and
  \item $(K{+}1)N{-}1$ additions ($K{+}1$ accumulators, each with $N{-}1$ additions plus $K$ additions of combining outputs).
\end{itemize}
Consequently, this approach avoids $N$ multiplications by $v[n]$ as well as the additional cost of generating $n^K$ (i.e., avoiding $(K{-}1){\times}N$ multiplications for direct evaluation or the memory required by a precomputed lookup table).
\begin{theorem}[Equivalence and Uniqueness]\label{theorem:unify}
Let $K \in \mathbb{Z}_{\ge 0}$ and $N \ge 1$. Then the representation in \eqref{eq:c_kA_k} is equivalent to
\begin{equation}
n^{K} = \sum_{k=1}^{K+1} c_k \binom{N-n+k-2}{k-1}, \label{eq:n_k}
\end{equation}
and the coefficients $\{c_k\}$ exist and are unique provided that $N \ge K{+}1$.
\end{theorem}

\begin{proof}\label{proof:equivalence} 
  Substituting $A_k[N-1]$ in \eqref{eq:accum} into~\eqref{eq:c_kA_k} yields
  \begin{align}
    \sum_{n=0}^{N-1} n^{K} v[n] &= \sum_{k=1}^{K+1} c_k \sum_{n=0}^{N-1} \binom{N-n+k-2}{k-1} v[n]. 
    \label{eq:main_identity}
  \end{align}
 The equality must hold for all possible $v[n]$. In particular, $v[n]{=}\delta[n-m]$ (shifted unit impulse sequence, a.k.a. Kronecker delta, which equals $1$ at $n=m$ and $0$ otherwise) gives
  \begin{align}
    m^{K} &= \sum_{k=1}^{K+1} c_k \binom{N-m+k-2}{k-1},
  \end{align}
  which is exactly \eqref{eq:n_k} for $n=m$, establishing the equivalence.

  Existence of the coefficients, $\{c_k\}$, follows because the basis $\{\binom{N-n+k-2}{k-1}\}_{k=1}^{K+1}$ in \eqref{eq:n_k} spans the space of polynomials in $n$ of degree at most $K$, and therefore $n^K$ can always be expressed as a linear combination of the basis functions.  

  If the identity \eqref{eq:n_k} is required to hold for all integers $n$, then uniqueness follows immediately: each basis function $\binom{N-n+k-2}{k-1}$ is a polynomial in $n$ of degree $k{-}1$ (see Footnote~\ref{foo:order}), and the $K+1$ basis functions then correspond to a system of $K+1$ equations with a lower–triangular matrix with nonzero diagonal entries $(-1)^{k{-}1}/(k{-}1)!$. These polynomials are therefore linearly independent and span the space of polynomials of degree at most $K$, so the representation of $n^K$ as their linear combination is unique. 

  When $n$ is restricted to the finite range $0,1,{\dots},N{-}1$, however, uniqueness of $\{c_k\}$ requires at least $K{{+}}1$ sample points, i.e., $N {\ge} K{+}1$, so that the $N{\times}(K{+}1)$ system has full column rank. Under this condition, the coefficients $c_k$ in \eqref{eq:n_k} [equivalently in \eqref{eq:c_kA_k}] are uniquely determined \cite{roman1992advanced, strang2009introduction}.
\end{proof}

\begin{remark}
  Uniqueness is unconditional in the polynomial sense, but for a signal of length $N$, it is guaranteed provided $N \ge K{+}1$. 
  Since the signal length $N$ in practice is typically much larger than the power $K$, this condition is always satisfied.
\end{remark}

\subsection{Derivation of $\{c_k\}$ in Closed Form via Stirling Numbers} \label{sec:prop_solution}
The solution for $c_k$ can be found by explicitly solving \eqref{eq:n_k}. To this end, let $n{=}N{-}x$, expanding $(N{-}x)^{K}$ gives
\begin{subequations}\label{eq:n_k3}
\begin{align}
  n^{K} &= (N{-}x)^{K} = \sum_{m=0}^{K} (-1)^m \binom{K}{m} N^{K-m} x^m. \label{eq:taylor}
  \intertext{By applying the identity $x^m \hspace{-2pt} {=} \sum_{j=0}^{m} \stirlingii{m}{j} (-1)^{m-j} x^{\overline{j}}$ from~\eqref{eq:stirling_expansion}, and recalling $x^{\overline{j}} = j! \binom{x + j - 1}{j}$ from~\eqref{eq:rising_ref}, yields} \nonumber \\[-22pt] 
 (N {-} x)^{K} &= \sum_{m=0}^{K} (-1)^m \binom{K}{m} N^{K-m} \nonumber \\[-0pt] 
  & \times \sum_{j=0}^{m}  \stirlingii{m}{j} (-1)^{m-j} j! \binom{x+j-1}{j}, \label{eq:n_k3a} \\[-2pt] 
  \intertext{where $\stirlingii{m}{j}$ denotes the Stirling numbers of the second kind. The summation runs over all integer pairs $(m,j)$ with $0 \leq j \leq m \leq K$. 
  This triangular region can equivalently be described by $0 \leq j \leq K$ and $j \leq m \leq K$, which allows us to rearrange\footnotemark{} the order of summation as} 
 (N {-} x)^{K} &= \sum_{j=0}^{K} j! \binom{x + j - 1}{j} \times \nonumber  \\ 
  &\times\sum_{m=j}^{K} (-1)^m \binom{K}{m} N^{K - m} (-1)^{m - j} \stirlingii{m}{j}.
\end{align}
\begin{align}
  \intertext{Further, reindexing to $k\gets j+1$ and substituting $x=N-n$ leads to}
  (N{-}[N{-}n])^K &= \sum_{k=1}^{K+1} \sum_{m=k-1}^{K} (-1)^m \binom{K}{m} N^{K-m} \nonumber \\
  \times (-1)^{m - (k-1)}\hspace{-12pt}&\hspace{12pt}(k{-}1)! \stirlingii{\hspace{-2pt}m\hspace{-2pt}}{\hspace{-2pt}k{-}1\hspace{-2pt}}\hspace{-1pt} \binom{\hspace{-1pt}(N {-} n) {+} (k{-}1) {-} 1\hspace{-1 pt}}{k{-}1\hspace{-3pt}}  \\
  \intertext{which after simplification gives} 
 n^K = \sum_{k=1}^{K{+}1} \hspace{- 5pt}&\hspace{ 5pt}({-}1)^{k{-}1} (k{-}1)! \hspace{-5 pt} \sum_{m=k-1}^{K} \hspace{-2pt}\binom{K}{m} N^{K{-}m} \nonumber \\
  \times \stirlingii{m}{k{-}1} \hspace{- 18pt}&\hspace{18pt} \binom{N{-}n{+}k{-}2}{k{-}1}. \label{eq:n_k3e}
\end{align}
\end{subequations}
\footnotetext{Note that 
$\displaystyle \sum_{m=0}^{K} \sum_{j=0}^{m} f(m, j){=}\hspace{-14 pt}\sum_{0\leq j\leq m\leq K}\hspace{-13 pt}f(m,j) {=} \sum_{j=0}^{K} \sum_{m=j}^{K} f(m, j)$.}

Thereafter, comparing \eqref{eq:n_k3e} with \eqref{eq:n_k}, directly leads to the expression
\begin{subequations}
  \begin{align}
      c_k &= (-1)^{k-1} (k-1)! \sum_{m=k-1}^{K} \binom{K}{m} N^{K-m} \stirlingii{m}{k{-}1}. \label{eq:ck_raw} \\
      \intertext{Further as $\stirlingii{m}{k-1}=0$ for $m<k-1$, the lower limit in \eqref{eq:ck_raw} can be set to $m=0$, which gives  
      }
      c_k &= (-1)^{k-1} (k-1)! \sum_{m=0}^{K} \binom{K}{m} N^{K-m} \stirlingii{m}{k{-}1}. \\
      \intertext{Then, by substituting the identity {$(-1)^{k-1}(k-1)!\,\stirlingii{m}{k-1}=\sum_{j=0}^{k-1}(-1)^j\binom{k-1}{j}\,j^m$} [see~\nameref{sec:app_identity_open}], yields}
      c_k &= \sum_{j=0}^{k-1} (-1)^j\binom{k{-}1}{j} \sum_{m=0}^{K} \binom{K}{m} N^{K-m} j^{m}.   
      \label{eq:ck_open}
  \end{align}
\end{subequations}
Lastly, applying the binomial theorem~\cite[eq.~(5.12)]{concrete_mathematics} leads to the simplified closed-form expression
\begin{align}
      c_k&= \sum_{j=0}^{k-1} (-1)^j \binom{k{-}1}{j} (N{+}j)^{K}, \quad k \in \hspace{-1 pt} \mathbb{Z}[1, K{+}1].  \label{eq:ck_closed}
\end{align}

The coefficients $\{c_k\}$ in \eqref{eq:ck_closed} are polynomials in $N$. Their degrees follow the pattern $K {-} (k {-} 1)$, meaning $c_{K+1}$ is a constant (degree 0) and $c_1$ has the highest degree, $K$. This arises from the telescoping property of the alternating binomial sum in~\eqref{eq:ck_closed} which acts as the $(k{-}1)$-th finite difference operator of the function $f(N) = (N{+}j)^K$ (see \cite[eq.~(5.40)]{concrete_mathematics}), which cancels all terms where $N$'s power exceeds $K {-} k {+} 1$, despite the $(N{+}j)^{K}$ term suggesting higher degrees. This pattern is verified in Table~\ref{tab:ck_varying_K} for $1 \leq k \leq K{+}1$ and $0 \leq K \leq 5$.

\begin{table*}[!ht]
  \centering
  \caption{Coefficients $\{c_{k}\}$ for $0 {\leq} K {\leq} 5$ and $1 {\leq} k {\leq} K{+}1$ \label{tab:ck_varying_K}}
  \begin{tabular}{c|c|c|c|c|c|c}
      $K$ &$c_1(N)$ &$c_2(N)$ &$c_3(N)$ &$c_4(N)$ &$c_5(N)$& $c_6(N)$\\
      \hline
      \hline
      $0$ & $1$ & \NA & \NA & \NA & \NA & \NA  \\
      $1$ & $N$ & $-1$ &\NA & \NA  &\NA  & \NA \\
      $2$ & $N^2$ & $-2N-1$ & $2$ &\NA & \NA & \NA  \\
      $3$ & $N^3$ & $-3N^2 - 3N - 1$ & $6N+6$ &$-6$ &\NA & \NA  \\
      $4$ & $N^4$ & $-4N^3 - 6N^2 - 4N - 1$ & $12N^2+24N+14$ &$-24N-36$ & 24 &  \NA \\
      $5$ & $N^5$ & $-5N^4 - 10N^3 -10N^2 -5N -1$ &$20N^3 + 60N^2 + 70N + 30$ &$ -60N^2 - 180N - 150$ & $120N+240$ & $-120$ \\
  \end{tabular}
\end{table*}

\begin{remark}
  Although $c_k$ are polynomials in $N$, the value of $N$ is typically known at runtime. 
  Hence the coefficients can be precomputed (or evaluated once at $n{=}N{-}1$), so that the hardware only requires $K{+}1$ constant multiplications $c_k A_k[N{-}1]$ in the implementation of the scheme in Fig.~\ref{flo:pipeline}. It is also noted from Table~\ref{tab:ck_varying_K} that selecting $N$ as some power of two can give particularly simple coefficients $c_k$, since $N^k$ then correspond to simple shift operations in hardware.
\end{remark}

\section{Complexity Analysis} \label{sec:results}  
This section compares the arithmetic complexity of evaluating the sum $\sum_{n=0}^{N-1} n^{K} v[n]$ when each term $n^{K}$ is computed at runtime through optimal addition-chain exponentiation\footnote{\label{foo:chain}An optimal addition-chain exponentiation is a sequence of multiplications that computes a power $n^{K}$ with the minimal number of multiplications. For example, only four multiplications are needed to compute $n^{7}$ via the chain \mbox{$n \xrightarrow{\times n} n^2 \xrightarrow{\times n} n^3 \xrightarrow{\times n^3} n^6 \xrightarrow{\times n} n^{7}$}.} (baseline), as described in~\cite{hatem2021}, and the proposed method, given by the right-hand side of~\eqref{eq:c_kA_k} with coefficients $c_k$ given by~\eqref{eq:ck_closed}, and implemented with the cascaded-accumulators structure in Fig.~\ref{flo:pipeline} that requires only $K{+}1$ constant multiplications after accumulation.\footnote{The code for evaluating the closed-form coefficients $c_k$ in \eqref{eq:ck_closed} and reproducing the numerical results of this letter is publicly available at: \url{https://gitlab.liu.se/spl/SPL-44561-2025}.}

\begin{figure}[t]
  \centering
  \includegraphics[scale=1]{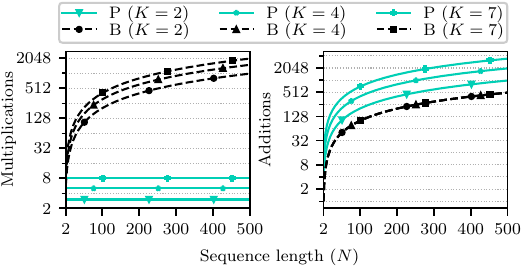}
  \caption{Arithmetic complexity. Proposed (P) against baseline (B) [optimal addition-chain exponentiation, see Footnote \ref{foo:chain}]. Vertical axis is logarithmic.}
  \label{flo:complexity}
\end{figure}

Figure~\ref{flo:complexity} presents both approaches for $K$ equal to $2$, $4$, and $7$. 
It shows that the multiplicative complexity of the baseline method grows linearly with the input size as~$M\times N$, where~$M$ is determined by the optimal addition-chain exponentiation, e.g., $M=4$ for computing $n^{7}$ (see Footnote~\ref{foo:chain}).  
In contrast, the proposed approach requires only \mbox{$K{+}1$} multiplications, independent of~$N$, at the expense of additional additions.\footnote{The approach in~\cite{Zhou2000} uses the same number of multiplications as the proposed method with different coefficients, but requires block-based processing with buffering of all $N$ input samples $v[n]$.} 
As mentioned earlier, since multiplications typically consume more power and area than additions~\cite{parhi2007vlsi, koren2001computer}, the proposal offers a substantial reduction in implementation complexity.
\subsection{Hardware Cost}
\label{sec:hw_cost}
To address hardware-level cost, a Verilog~RTL implementation of the proposed architecture and a straightforward baseline were synthesized using Yosys~0.53 (\texttt{synth\_ecp5}) and subsequently packed and placed-and-routed using \texttt{nextpnr}/\texttt{ecppack} targeting a Lattice ECP5 device.
Synthesis was performed for a representative configuration ($K{=}2$, $N{=}100$), where $v[n]$ contain the first $N{=}100$ semiprime numbers~\cite{OEIS_A001358}, implying no quantization error in a 32-bit implementation. Exponentiation of $n^{K}$ is assumed precomputed and stored for the baseline. Its exponentiation and memory-access costs are not included, thereby favoring the baseline.
\begin{table}[t]
\centering
\caption{Resource utilization (pre-pack and post place-and-route).}
\label{tab:hw_cost}
\begin{tabular}{lrrr}
\toprule
\textbf{Metric} & \textbf{Baseline} & \textbf{Proposed} & \textbf{Change} \\
\midrule
Latency (cycles, RTL) & 100 & 102 & $+2\%$ \\
LUT4s before packing (\texttt{LUT4}) & 152 & 96 & $-37\%$ \\
\midrule
Logic cells (\texttt{TRELLIS\_COMB}) & 182 & 110 & $-40\%$ \\
Flip-flops (\texttt{TRELLIS\_FF}) & 96 & 96 & -- \\
DSP blocks (\texttt{MULT18X18D}) & 6 & 0 & $-100\%$ \\
Fixed-point accuracy (32-bit)$^{\dagger}$ & Exact & Exact & -- \\
\bottomrule
\end{tabular}
\vspace{1mm}
\begin{minipage}{\linewidth}
\footnotesize\raggedright
$^{\dagger}$Reducing arithmetic precision would further favor the proposed method, as quantization error accumulates with the number of multiplications and the proposed method performs only $K \ll N$ multiplications per block.
\end{minipage}
\vspace{-12 pt}
\end{table}
As seen in Table~\ref{tab:hw_cost}, the proposed architecture reduces the overall resource utilization for this example.
\section{Conclusion}\label{sec:conclusion}
\vspace{-1 pt}
This work presented an efficient method for computing time-index powered weighted sums using cascaded accumulators. 
The key contribution is the decomposition of \eqref{eq:S_k} into a linear combination of accumulator outputs [see~\eqref{eq:c_kA_k}], which enables efficient real-time sample-by-sample computation. The proposed architecture eliminates the need for storing entire data blocks, making it well suited for streaming, low-power, and resource-constrained signal processing applications. 
\vspace{-1 pt}
\appendix{Identity used to obtain \eqref{eq:ck_open}}
\label{sec:app_identity_open}
Starting with the Stirling number of the second kind $\stirlingii{m}{k-1}$ [See \eqref{eq:stirling_second}] and multiplying both sides by $(-1)^{k-1}(k-1)!$ gives
\begin{align*}
  (\hspace{-2pt}{-}1)^{k{-}1}\hspace{-1pt}(\hspace{-1pt}k{-}1)! \stirlingii{m}{k{-}1} {=} \hspace{-2pt}\sum_{i=0}^{k{-}1} ({-}1)^{i{+}k{-}1} \binom{k{-}1}{i} (k{-}1{-}i)^m, \\
  \intertext{making the substitution $j {=} k{-}1{-}i$ ($\Rightarrow i {=} k{-}1{-}j$)}
  {=} \hspace{-7pt}\sum_{j=k{-}1}^0 \hspace{-5pt}({-}1)^{(k{-}1{-}j){+}(k{-}1)} \binom{k{-}1}{k{-}1{-}j} j^m 
  {=} \hspace{-2pt}\sum_{j=0}^{k{-}1} ({-}1)^{j} \binom{k{-}1}{j} j^m.
\end{align*}

\begin{small} 
	\bibliographystyle{IEEEtran}
   \bibliography{IEEEabrv, references/bibliography} 
\end{small}

\end{document}